\documentclass[trackchanges]{aastex701}

\usepackage{graphicx}
\usepackage[version=4]{mhchem}
\usepackage{csquotes}

\begin{document}

\title{Nautilus Space Observatory: The Evolution of Planets and their Atmospheres}

\author[orcid=0000-0001-7962-1683]{Ilaria Pascucci}
\affiliation{Lunar and Planetary Laboratory, University of Arizona, 1629 E. University Boulevard, Tucson, AZ 85721, USA}
\affiliation{Steward Observatory, The University of Arizona, 933 N. Cherry Avenue, Tucson, AZ 85721, USA}
\email[show]{pascucci@arizona.edu} 

\author[orcid=0000-0003-3989-5545]{Noah Tuchow}
\affiliation{Lunar and Planetary Laboratory, University of Arizona, 1629 E. University Boulevard, Tucson, AZ 85721, USA}\email{nwtuchow@arizona.edu}

\author[orcid=0000-0003-2969-6040,sname='Zhou']{Yifan Zhou}
\affiliation{Department of Astronomy, University of Virginia, 530 McCormick Rd., Charlottesville, Virginia, 22904 USA}
\email{yzhou@virginia.edu}

\author[orcid=0000-0003-3714-5855]{D\'aniel Apai}
\affiliation{Steward Observatory, The University of Arizona, 933 N. Cherry Avenue, Tucson, AZ 85721, USA}
\affiliation{Lunar and Planetary Laboratory, University of Arizona, 1629 E. University Boulevard, Tucson, AZ 85721, USA}
\affiliation{Alien Earths Team, NASA ICAR/NExSS, USA}
\affiliation{Department of Earth, Atmospheric and Planetary Science, MIT, 77 Massachusetts Avenue, Cambridge, MA 02139, USA} \email{apai@arizona.edu}

\author[orcid=0009-0007-4584-4417]{Chaucer Langbert}
\affiliation{Lunar and Planetary Laboratory, University of Arizona, 1629 E. University Boulevard, Tucson, AZ 85721, USA}
\email[]{} 

\author[orcid=0000-0002-5322-2315]{Ana Glidden}
\affiliation{Department of Earth, Atmospheric and Planetary Science, MIT, 77 Massachusetts Avenue, Cambridge, MA 02139, USA} \email[]{}

\author[orcid=0000-0003-0156-4564]{Luis Welbanks}
\affiliation{School of Earth and Space Exploration, Arizona State University, Tempe, AZ, USA}
\email[]{luis.welbanks@asu.edu}

\author[orcid=0000-0001-5989-7594]{Chia-Lung Lin}
\affiliation{Steward Observatory, The University of Arizona, 933 N. Cherry Avenue, Tucson, AZ 85721, USA}
\email{chialunglin@arizona.edu}

\author[orcid=0000-0002-9464-8101]{Adina~D.~Feinstein}
\affiliation{Department of Physics and Astronomy, Michigan State University, East Lansing, MI 48824 USA}
\email[]{adina@msu.edu} 

\author[orcid=0000-0002-3627-1676]{Benjamin V. Rackham}
\affiliation{Department of Earth, Atmospheric and Planetary Science, MIT, 77 Massachusetts Avenue, Cambridge, MA 02139, USA}
\affiliation{Kavli Institute for Astrophysics and Space Research, Massachusetts Institute of Technology, Cambridge, MA 02139, USA}
\email[]{brackham@mit.edu}

\author[orcid=0000-0002-8864-1667]{Peter Plavchan}
\affiliation{Department of Physics and Astronomy, George Mason University, Fairfax, VA, USA} \email[]{}

\author[0000-0002-4309-6343]{Kevin Wagner}
\affiliation{Department of Astronomy and Steward Observatory, University of Arizona, Tucson, AZ 85721, USA}
\email[]{kevinwagner@arizona.edu}

\author[orcid=0000-0002-5887-1197] {Raymond Pierrehumbert}
\affiliation{University of Oxford, Department of Physics Oxford, OX1 3PW, UK} \email[]{}  

\author[orcid=0000-0003-1127-8334]{Robin Wordsworth}
\affiliation{Department of Earth and Planetary Sciences, Harvard University, Cambridge, MA 02138, USA}
\email[]{}  


\begin{abstract}
We are just beginning to explore the billion-year evolution from nascent planets in disks to mature planetary systems.
Recent discoveries hint at demographic and atmospheric differences between young planets and their Gyr-old counterparts, but current facilities are limited $-$ particularly in their ability to conduct statistical atmospheric studies over a broad period range. This white paper outlines compelling science achievable with the Nautilus Space Observatory, a proposed constellation of large-diameter space telescopes. We identify four primary scientific objectives: (1) determining the timescales over which planets evolve into sub-Neptunes and super-Earths; (2) tracking the temporal evolution of atmospheric mass-loss rates; (3) characterizing the evolution of the atmospheric mean molecular weight and C/O ratio; and (4) identifying the emergence of Helium-dominated worlds. Answering these questions requires the high spatial resolution, broad-wavelength coverage, large effective area, and parallelized multiple units that Nautilus provides. By isolating the physical processes that govern the evolution of planets and their atmospheres, these science objectives directly support NASA's Cosmic Origins and Exoplanet Exploration Programs.
\end{abstract}



\section{Scientific Context and Problem Statement} \label{sect:intro}
This White Paper presents a potential science case for the Nautilus Space Observatory, a concept under development for a NASA Strategic Mission for the Astro 2030 Decadal Survey. Nautilus is a constellation of space telescopes and will provide a modular, scalable, sustainable, upgradable, expandable space observatory that can be deployed rapidly and then expanded progressively. The core concept for Nautilus is described in \cite{Apai2019AJ....158...83A}.  This White Paper is part of the first series of science white papers capturing ideas that emerged from the Nautilus Science Case workshop (held at MIT in May 2026). This White Paper focuses on the evolution of planets and their atmospheres.

JWST and ALMA are providing unprecedented detail on the gas and dust of hundreds of protoplanetary disks during the first few tens of millions of years of planetary assembly \citep[e.g.,][]{Pascucci2016ApJ...831..125P,Andrews2018ApJ...869L..41A,Long2018ApJ...869...17L,Coco2025ApJ...989....1Z}.
In parallel, exoplanet surveys, relying primarily on the transit and radial velocity methods, have discovered over 6,000 exoplanets \citep[e.g.,][]{Rose2021ApJS..255....8R,Kunimoto2022ApJS..259...33K,Lissauer2024PSJ.....5..152L}. However, the vast majority of these are Gyr-old planets and planetary systems. A billion years of evolution from nascent planets to mature systems is only now beginning to be explored.

Two dozen young ($\sim 20-1,000$ Myr) Neptunes and sub-Neptunes recently discovered by K2 and TESS are providing the first evidence of demographic changes between nascent and mature systems. It is found that their frequency,  within 10\,days, is higher than that of their Gyr-old counterparts \citep{Ch2023AJ....166..248C,Vach2024AJ....167..210V}. This decline does not appear to be linear over time \citep{Fernandes2025AJ....169..208F}, which has been attributed to a combination of tidal migration and photoevaporation, though small-number statistics preclude firmly establishing this trend. In addition, the fraction of neighboring planets near a first-order orbital resonance is larger for the dozen of known younger than older systems \citep{Dai2024AJ....168..239D}.
Finally, there is emerging evidence for atmospheric evolution. Recent HST and JWST spectroscopic observations of a few young planets indicate that they have clearer atmospheres and lower metallicities ($\sim$solar to a few times solar), hence lower mean molecular weight, than older planets of comparable mass \citep{Thao2024AJ....168..297T,Barat2025AJ....170..165B}.

Isolating the physical processes that drive the evolution of planets and their atmospheres requires detecting and characterizing a larger sample of young planets across a broader range of orbital periods. Current facilities cannot achieve this: TESS searches are restricted to nearby star-forming regions ($\leq 200$\,pc) due to sensitivity and severe flux contamination from its large pixels \citep[e.g.,][]{Fernandes2022AJ....164...78F};  JWST characterization is limited by the sample of discovered young planets, mostly within 10 days, and high proposal oversubscription; and any Roman's spectroscopic characterization is restricted to a narrow grism window (1.0–1.9$\,\mu$m). Nautilus uniquely overcomes these bottlenecks.
By combining high spatial resolution and broad-wavelength spectroscopy, with multiple units for efficient mapping of star-forming regions, Nautilus is the ideal facility to conduct comprehensive demographic studies of young planets and their atmospheres.





\section{Science Objectives} 
\label{sect:so}
\begin{figure*}[ht!]
\centering
\includegraphics[width=\textwidth, height=0.5\textheight, keepaspectratio]{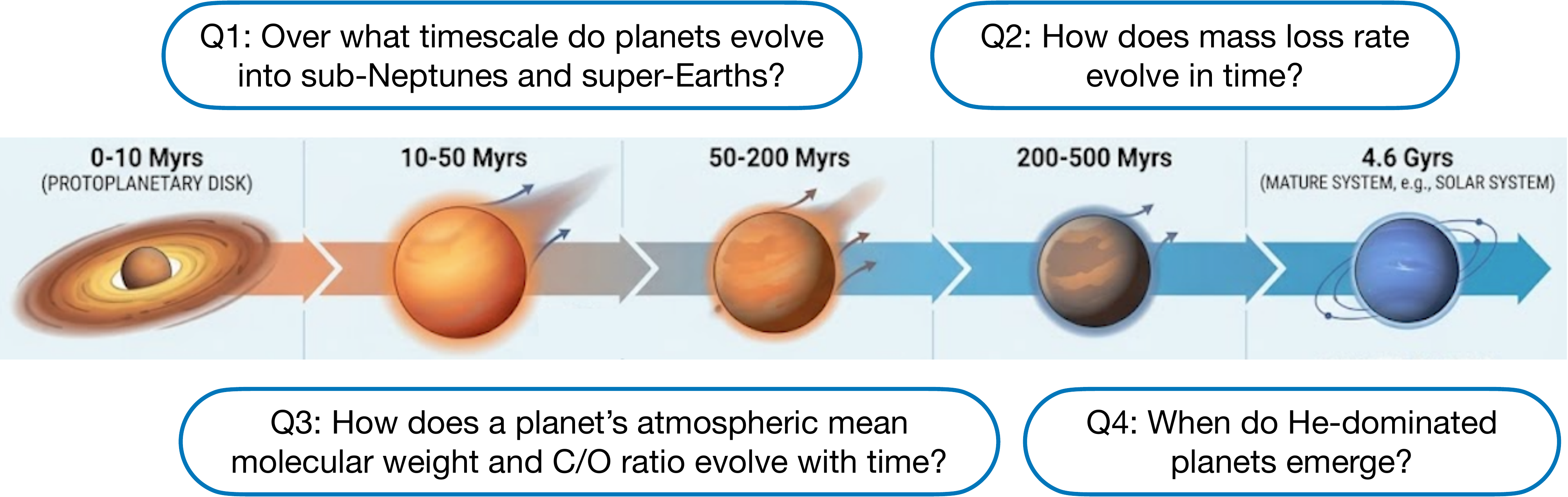}
\caption{Figure summarizing the key science objectives.}
\label{fig:questions}
\end{figure*}

The demographics of young exoplanets, combined with the characterization of their atmospheres through transmission spectroscopy, will enable Nautilus to answer the following questions (see Fig.~\ref{fig:questions} for an overview):
\begin{enumerate}
\item \textbf{Over what timescale do planets evolve into sub-Neptunes and super-Earths?}
Sub-Neptunes ($\sim 1.8-4$\,R$_\oplus$) and super-Earths ($\sim 1.2-1.8$\,R$_\oplus$) are the most common transiting exoplanets \citep[e.g.,][]{Lissauer2023ASPC..534..839L}, yet we have no equivalent in the Solar System. Understanding which planets enter these classes, and when, provides crucial context for the origins of planetary systems. Answering this question will require the measurement of occurrence rates of planets around young ($\sim 50-500$\,Myr) stars,  with radii down to at least $\sim 1.5$\,R$_\oplus$ and periods extending to at least 50\,days. The largest period is set by the transition period beyond which sub-Neptunes become more frequent than super-Earths around Gyr-old stars \citep{Bergsten2022AJ....164..190B}. 

\item \textbf{How does mass loss rate evolve in time?}
Atmospheric evaporation is one of the key processes invoked to explain the division between Gyr-old, mostly rocky super-Earths and larger, gas-rich or water-world sub-Neptunes \citep{Fulton2017AJ....154..109F}. Pinning down the evolution of the mass-loss rate will tell us whether most evaporation is driven by high-energy stellar photons early on or by core-powered mass loss at later times \citep[e.g.,][]{OW2017ApJ...847...29O,Ginzburg2018MNRAS.476..759G}. It will also help clarify whether other mechanisms, e.g., formation at different locations in the disk \citep[e.g.,][]{Burn2024NatAs...8..463B}, are also critical in setting the separation of these two classes of planets. The He~I line at 1.083~$\mu$m is a reliable tracer of atmospheric escape \citep[e.g.,][]{GS2024AJ....167..142G}. Even when not spectrally resolved, it can provide a first-order assessment of the mass-loss rate, especially when coupled with models that compute the wind thermal structure \citep{Linssen2022A&A...667A..54L}. When spectrally resolved (requiring $R \ge 30,000$) and combined with other hydrogen lines, it provides more stringent constraints on the mass-loss rate \citep[e.g.,][]{Lampon2021A&A...647A.129L}.

\item \textbf{How do a planet's atmospheric mean molecular weight and C/O ratio evolve with time?}  Billions of years of magma ocean outgassing and atmospheric escape severely contaminate the atmospheric compositions of Gyr-old planets \citep[e.g.,][]{Kite2020PNAS..11718264K,Nicholls2024JGRE..12908576N}. 
This erases signatures of their initial formation location in the disk and, for sub-Neptunes, their bulk composition (whether they are gas dwarfs of water worlds). Measuring  the C/O ratio and mean molecular weight of young planets could reveal the formation locations and the nature of sub-Neptunes \citep[e.g.,][]{Madu2016SSRv..205..285M,Rogers2025MNRAS.539.2230R}.
 For this science goal, the minimum requirement is to measure water abundances (even at moderate resolution $R \sim 100$) to infer the scale height and mean molecular weight \citep[e.g.][]{BS2013ApJ...778..153B}, while detecting CH$_4$ and   CO$_2$ is necessary to estimate the C/O ratio \citep[e.g.,][]{Bat2017AJ....153..151B}.

\item \textbf{When do He-dominated planets emerge?}
Helium worlds are a robust prediction of atmospheric escape models, but they represent a transient state that lasts on the order of a hundred million years \citep[e.g.,][]{Hu2015ApJ...807....8H,Cherubim2025ApJ...983...97C}. They are expected to be located below the Neptune desert ($\sim 3-10$\,R$_\oplus$, $< 4$\,days) and at the upper edge of the radius valley \citep{Cherubim2024ApJ...967..139C}. Thus, their detection is important for understanding hydrodynamic atmospheric escape, as well as the origin of the Neptune desert and radius valley. These planets may be identified through the detection of He and CO$_2$ \citep[e.g.,][]{Barkaoui2024NatAs...8..909B}, which are expected to be abundant, alongside constraints on CH$_4$, which is expected to be significantly depleted \citep[e.g.,][]{Cherubim2025ApJ...983...97C}.
\end{enumerate}


\section{Data Requirements}
Of the science objectives outlined above, only the first can be achieved with broad-band photometry while the others require transmission spectroscopy. 
\begin{itemize}
\item \textbf{Photometry:} Multiple filters ($\sim 3$) are required for photometric transit searches of young exoplanets and for validating detections. A 2-10\,minute cadence is sufficient for this science case (but note that with 20-30\,second cadence science questions related to stellar flares can be also addressed). To  detect planets out to 50\,days we require continuous monitoring for at least 150\,days. A photometric precision of 15\,ppm is needed to detect a super-Earth (=1.5\,R$_\oplus$) with SNR of 10 over 3.5\,h (half of the transit duration for a planet at 50\,days orbiting a Sun-like star).
An angular resolution of $\sim 0.1''$ will expand the sample of accessible young stars by more than an order of magnitude and enable typical binary systems ($\sim 40$ au) to be resolved out to 400 pc.
\item \textbf{Spectroscopy}. The second science question requires modest spectroscopic capabilities covering from $\sim 0.4-1.5$\,$\mu$m at (R$\sim 500$).
High-precision optical and near-infrared spectra are needed to derive joint constraints on cloud/haze properties and chemical abundance \citep[e.g.,][]{Welbanks2019AJ....157..206W} and detect the He\,I line at 1.083\,\micron. Resolving the line (typical width $\sim 20$\,km/s) will require higher spectral resolution ($R \ge 30,000$). Science objectives 3 and 4 can be achieved at lower spectral resolution ($R \sim 1,000$). Water lines are present even in the short wavelength range mentioned above and are sufficient to address the evolution of a planet's mean molecular weight. Detecting CO$_2$ and CH$_4$ (relevant for the C/O ratio and the identification of He worlds) requires extending the wavelength coverage to at least 2.5\,$\mu$m. 
\end{itemize}


\section{Analysis and Interpretation} 
The necessary datasets are  time-averaged photometry for Q1 and time-averaged spectra for Q2–Q4. Well established transit analysis scripts $-$ originally developed for Kepler, TESS, and JWST $-$ can be utilized to analyze both datasets. Despite the large photometric variability of young stars ($\sim 10,000$\,ppm), modern detrending techniques have successfully pushed detection limits down to the super-Earth regime \citep[e.g.,][]{Mann2018AJ....155....4M, Barber2025AJ....170...32B, Fernandes2025AJ....169..208F}.

\section{Relevant Science Requirements}
System-level science requirements are summarized in Tables~\ref{scireq1} and ~\ref{scireq2}.


\section{Relevance to Nautilus and Mission Class} 
As mentioned in Sect.~\ref{sect:intro}, current facilities face significant bottlenecks in characterizing a large sample of young planets across a broad range of orbital periods—limitations that Nautilus can overcome. Here, we further stress that because star-forming regions and associations are distributed across vast sky areas $-$ often spanning tens of square degrees $-$ a constellation of telescopes with the parallelized observing capabilities of Nautilus becomes essential for accelerating survey execution.




\begin{deluxetable}{lccl}
\tabletypesize{\scriptsize}
\tablecaption{Summary of system-level science requirements and associated  objectives as listed in Sect.~\ref{sect:so}. \label{scireq1}}
\tablewidth{0pt}
\tablehead{
\colhead{Requirement} & \colhead{Imaging} & \colhead{Spect.} & \colhead{Science Driver} 
}
\startdata
250-350\,nm  & N/A  & Optional & Q2+Q3 (Rayleigh scattering) \\
350-450\,nm  & Optional & Optional &  Q2+Q3 (Rayleigh scattering, stellar variability)\\
450-1,000\,nm  & Required & Required & Q1 (exoplanet photometric detection in multiple bands, Rayleigh scattering) \\
1-1.8$\mu$m  & Required & Required & Q1 (J or H, reduced stellar variability) + Q2 (He\,I at 1.083\,$\mu$m) + Q3 (water, several lines) \\
1.8-2.3$\mu$m  & N/A &  Required & Q3 (CO$_2$ at $\sim 2$\,$\mu$m)\\
2.3-2.9$\mu$m  & N/A & Required  & Q3+Q4 (CH$_4$ at $\sim 2.3$\,$\mu$m, CO$_2$ at $\sim 2.8$\,$\mu$m)\\
\enddata
\end{deluxetable}

\begin{deluxetable}{lll}
\tabletypesize{\scriptsize}
\tablecaption{Necessary and optional system-level science requirements.\label{scireq2}}
\tablewidth{0pt}
\tablehead{
\colhead{Requirement} & \colhead{Range} & \colhead{Science Driver}  \\
}
\startdata
Photometric Filters & VIJ & Q1 (multi-band for exoplanet  validation, at J stellar variability is reduced)\\
Target Brightness [mag] & 8-27 & to cover stars from G to M and distances from 50 to 500\,pc   \\
Min. Photom. Precision [ppm] & 15 & over 3.5\,h to detect a super-Earth around a Sun-like star out to 50\,days   \\
Image Res. [diff. limit] & 0.1 & to separate a typical binary out to 400\,pc  \\
Min. Sky Coverage [deg$^2$] & a few & to efficiently map the tens of deg$^2$ of star-forming regions and association  \\
Min. Contrast & N/A& \\
Spectral Resolving Power & 500-30,000  & 30,000 is set to spectrally resolve the He I (part of Q2), all others require 500-1,000 \\
\hline
Relevant Timescales [h] & 3-7  & max transit durations for planets at 5-50\,days \\
Monitoring Baseline [d] & 150 & to detect transits out to 50\,days \\
Cadence [s] & 120-600 & comparable to TESS  \\
Rapid Response Time [s] & N/A & \\
\hline
Data Volume &   & \\
Pointing Precision [arcsec] & $<1/5$th of a pixel  & to reduce jitter and extract precise photometry and spectroscopy \\
\enddata
\end{deluxetable}
\section{Relevance to NASA and Astrophysics Strategy} 
Characterizing young exoplanetary systems directly supports NASA’s Cosmic Origins key question \enquote{How did we get here?} (Strategy 1.1 of NASA's Science Plan\footnote{https://assets.science.nasa.gov/content/dam/science/cds/about-us/2025/2025-2026-NASA-Science-Plan$\_$Tagged.pdf}). The physical evolution of smaller planets like super-Earths can also impact their habitability, inherently linking this research to a question central to NASA's Exoplanet Exploration, \enquote{Are we alone?}. As such, the study of young exoplanets acts as a critical bridge between Cosmic Origins and Exoplanet Exploration, two NASA astrophysics science themes, and directly advances the Astro2020 Decadal Survey's priority to understand \enquote{Worlds and Suns in Context} \citep{Astro2020_2021pdaa.book.....N}.


\begin{acknowledgments}
We thank the Heising-Simons Foundation for supporting the Nautilus Science Case Workshop. 
\end{acknowledgments}

\bibliography{sample701}{}
\bibliographystyle{aasjournalv7}

\end{document}